\documentclass[a4paper,12pt]{article}

\usepackage[hmargin=.71in,vmargin=1.1in]{geometry}
\usepackage{indentfirst}
\usepackage{lmodern}
\usepackage{amsfonts}
\usepackage{subfig}
\usepackage{mathrsfs}
\usepackage{amsmath}
\usepackage{mathabx}
\usepackage{amssymb}
\usepackage{authblk}
\usepackage{cite}
\usepackage{xcolor}
\usepackage{mathtools}
\usepackage{tensor}
\usepackage{physics}
\usepackage{graphicx}
\usepackage{bm}
\usepackage{upgreek}
\usepackage{braket}
\usepackage{color,soul}
\usepackage{csquotes}
\usepackage{caption}
\usepackage{slashed}
\usepackage{mathtools}
\usepackage{multirow}
\usepackage{listings}

\usepackage[bookmarksnumbered=true,bookmarksopen=true]{hyperref}
 \hypersetup{colorlinks,
             linkcolor=[rgb]{0,0.3,0.6}, 
             citecolor=[rgb]{0,0.3,0.6}, 
             urlcolor=[rgb]{0,0.3,0.6}}

\definecolor{verylightgrey}{rgb}{0.96,0.96,0.96}
\lstdefinestyle{mylststyle}{
    backgroundcolor=\color{verylightgrey},   
    columns=flexible,
    commentstyle=\color{green},
    keywordstyle=\bfseries\color{blue},
    basicstyle=\ttfamily\footnotesize,
    breakatwhitespace=false,         
    breaklines=true,                 
    captionpos=b,                    
    keepspaces=true,                 
    numbers=left,                    
    numbersep=5pt,                  
    showspaces=false,                
    showstringspaces=false,
    showtabs=false,                  
    tabsize=2
}
\renewcommand{\d}{\mathrm{d}}

\newcommand{\romannumup}[1]{\uppercase\expandafter{\romannumeral#1}}
\newcommand{\romannumlow}[1]{\expandafter{\romannumeral#1}}

\usepackage{tikz}
\usetikzlibrary{matrix}

\begin{document}

\title{Comment on ``Black holes in $f(\mathbb{Q})$ gravity''}

\author[a]{Zhen-Xiao Zhang\thanks{zx.zhang@mail.nankai.edu.cn}}	
\author[b]{Chen Lan\thanks{stlanchen@126.com}}
\author[a]{Yan-Gang Miao\thanks{Corresponding author: miaoyg@nankai.edu.cn}}

\affil[a]{\normalsize{\em School of Physics, Nankai University, 94 Weijin Road, Tianjin 300071, China}}
\affil[b]{\normalsize{\em Department of Physics, Yantai University, 30 Qingquan Road, Yantai 264005, China}}

\date{ }
\maketitle

\begin{abstract}
In the work [\textit{Phys.\ Rev.\ D 105 (2022) 2, 024042}], D'Ambrosio \textit{et al.}\ investigated spherically symmetric black hole solutions in $f(\mathbb{Q})$ gravity, where several solutions satisfy the condition: $g_{tt}g_{rr} = \mathrm{const}$. This condition is characteristic of many black holes, including the Schwarzschild spacetime. In this Comment, we argue that no nontrivial vacuum black hole solutions satisfy this condition in $f(\mathbb{Q})$ gravity. We demonstrate our claim by reexamining the field equations under the ``Set 2'' connection called by D'Ambrosio \textit{et al.}, which is necessary for obtaining solutions distinct from those of general relativity (GR). For the case where the free parameters $c$ and $k$ are zero, i.e.,  Option 2 in their work, we show that any attempt to find a solution beyond GR forces the non-metricity scalar to vanish ($\mathbb{Q}=0$), which trivializes the field equations and does not describe a valid black hole solution. Our findings indicate that the condition, $g_{tt}g_{rr} = \mathrm{const.}$, is overly restrictive for finding new, static and spherically symmetric vacuum black holes in $f(\mathbb{Q})$ gravity. This conclusion does not depend on the specific form of $f(\mathbb{Q})$. We also briefly discuss Option 1 that was not addressed in D'Ambrosio \textit{et al.}'s work, and give new constraints for the selection of parameters $c$ and $k$.
\end{abstract}

In the work~\cite{d2022black}, D'Ambrosio \textit{et al.} proposed their methods for solving spherically symmetric black holes in $f(\mathbb{Q})$ gravity and  deriving some approximate solutions, such as the most basic spherically symmetric vacuum solution under the quadratic $f(\mathbb{Q})$, $f(\mathbb{Q})=\mathbb{Q}+\alpha\mathbb{Q}^2$.
Many of these metrics satisfy the condition: $g_{tt}g_{rr}=\mathrm{const}$. This is a common condition in spherically symmetric black holes, indicating that the spacetime metric belongs to the Segre classification type $[(1,1)(11)]$~\cite{Stephani:2003tm}. However, we argue that there are no new vacuum solutions in $f(\mathbb{Q})$ gravity that satisfy the condition, $g_{tt}g_{rr}=\mathrm{const.}$, within the scope of the discussion in Ref.~\cite{d2022black}.


Since the connection $\tensor{\Gamma}{^\alpha_\beta_\gamma}$ is independent of the metric $\tensor{g}{_\alpha_\beta}$ and has its own field equation in $f(\mathbb{Q})$ gravity, which differs from the case in GR, it is necessary to simplify the connection as much as possible before solving it together with the metric. 

In D'Ambrosio \textit{et al.}'s work, the solutions of the connection are divided into two sets, listed in Table 1 and Table 2 of Ref.\ \cite{d2022black}, where Set 1 is proved to be equivalent to the solutions of GR. So, in order to discuss black hole solutions beyond GR, we need only to consider the black hole solutions in Set 2, where four independent components, $\tensor{\Gamma}{^t_r_r}(r)$, $\tensor{\Gamma}{^t_\theta_\theta}(r)$, $\tensor{\Gamma}{^r_r_r}(r)$, and $\tensor{\Gamma}{^r_\theta_\theta}(r)$, and two free parameters, $c$ and $k$, are enough to express the other components of the connection $\tensor{\Gamma}{^\alpha_\beta_\gamma}$ under the spherical symmetry and vanishing curvature and torsion. Therefore, a spherically symmetric connection solution can be represented by
\begin{equation}
    \{c, k, \tensor{\Gamma}{^t_r_r}(r), \tensor{\Gamma}{^t_\theta_\theta}(r), \tensor{\Gamma}{^r_r_r}(r), \tensor{\Gamma}{^r_\theta_\theta}(r)\}.
\end{equation}
D'Ambrosio \textit{et al.} also derived two differential equations,
\begin{subequations}
    \begin{equation}
        \partial_r\tensor{\Gamma}{^t_\theta_\theta}=-\frac{\tensor{\Gamma}{^t_\theta_\theta}}{\tensor{\Gamma}{^r_\theta_\theta}}[1+\tensor{\Gamma}{^t_\theta_\theta}(3c-k+(2c-k)\tensor{\Gamma}{^t_\theta_\theta})]-\tensor{\Gamma}{^t_r_r}\tensor{\Gamma}{^r_\theta_\theta},
        \label{eq:oridif1}
    \end{equation}
    \begin{equation}
        \partial_r\tensor{\Gamma}{^r_\theta_\theta}=-1-c\tensor{\Gamma}{^t_\theta_\theta}(2+(2c-k)\tensor{\Gamma}{^t_\theta_\theta})-\tensor{\Gamma}{^r_r_r}\tensor{\Gamma}{^r_\theta_\theta},
        \label{eq:oridif2}
    \end{equation}
\end{subequations}
and proved that one of the following two options must be satisfied,
\begin{align}
    &\mbox{Option 1:}\quad \begin{cases} \label{eq:option1}
        \tensor{\Gamma}{^t_\theta_\theta}=\frac{k}{2c(2c-k)},\\
        \tensor{\Gamma}{^t_r_r}=-\frac{k(8c^2+2ck-k^2)}{8c^2(2c-k)^2(\tensor{\Gamma}{^r_\theta_\theta})^2},\\
    \end{cases}
    \quad c\neq 0, \;\;\; k\neq 2c,\\
    &\mbox{Option 2:}\quad \begin{cases} \label{eq:option2}
        c=k=0,\\
        \tensor{\Gamma}{^t_r_r}=-\frac{\tensor{\Gamma}{^t_\theta_\theta}}{(\tensor{\Gamma}{^r_\theta_\theta})^2}.\\
    \end{cases}
\end{align}

Now we reexamine Option 2 in detail. Suppose the vacuum action takes the form,
\begin{equation}
    S=\frac{1}{2\kappa}\int \d^4 x \sqrt{-g}f(\mathbb{Q}),
\end{equation}
and it gives the following line element,
\begin{equation}
    \d s^2=-F(r)\d t^2+\frac{p}{F(r)}\d r^2+r^2\left(\d\theta^2+\sin^2\theta\d\varphi^2\right),
\end{equation}
where $p$ is constant ensuring $g_{tt}g_{rr} =-p= \mathrm{const.}$
Substituting the above ansatz into the vacuum metric field equation of $f(\mathbb{Q})$ gravity \cite{heisenberg2024review}, we obtain the two equations for the $tt$ and $rr$ components,
\begin{subequations}
    \begin{equation}
        -\frac{2\left(r F'(r)+F(r)-p\right)f'(\mathbb{Q})}{p r^2}+\frac{\mathbb{Q}'f''(\mathbb{Q})  }{p}\left[-\frac{2 F(r)}{\tensor{\Gamma}{^r_\theta_\theta}-\frac{4 F(r)}{r}}-\frac{2 p \tensor{\Gamma}{^r_\theta_\theta}}{r^2}\right]-\mathbb{Q}f'(\mathbb{Q}) +f(\mathbb{Q})=0,
        \label{eq:1_mfe1}
    \end{equation}
    \begin{multline}
        F(r) \tensor{\Gamma}{^r_\theta_\theta} \left[\left(2 r F'(r)+2 F(r)+p r^2 \mathbb{Q}-2 p\right)f'(\mathbb{Q})-p r^2 f(\mathbb{Q})\right]\\-\left(2 r^2 F^2(r)-2 p F(r) (\tensor{\Gamma}{^r_\theta_\theta})^2\right) \mathbb{Q}'f''(\mathbb{Q})=0.
        \label{eq:1_mfe2}
    \end{multline}
\end{subequations}
From Eqs.~\eqref{eq:1_mfe1} and \eqref{eq:1_mfe2}, we can solve for $F'(r)$,  respectively,
\begin{subequations}
    \begin{multline}
        F'(r)=\frac{-  \left[2 F(r)+p \left(r^2 \mathbb{Q}-2\right)\right]\tensor{\Gamma}{^r_\theta_\theta}f'(\mathbb{Q})}{2 r  \tensor{\Gamma}{^r_\theta_\theta}f'(\mathbb{Q})}\\
        +\frac{-2 \left[r F(r) (2 \tensor{\Gamma}{^r_\theta_\theta}+r)+p (\tensor{\Gamma}{^r_\theta_\theta})^2\right] \mathbb{Q}'f''(\mathbb{Q})+p r^2  \tensor{\Gamma}{^r_\theta_\theta}f(\mathbb{Q})}{2 r  \tensor{\Gamma}{^r_\theta_\theta}f'(\mathbb{Q})},
        \label{eq:2.partialF.1}
    \end{multline}
    \begin{equation}
        F'(r)=\frac{-\left[2 F(r)+p \left(r^2 \mathbb{Q}-2\right)\right]\tensor{\Gamma}{^r_\theta_\theta}f'(\mathbb{Q})+2  \left[r^2 F(r)-p (\tensor{\Gamma}{^r_\theta_\theta})^2\right] \mathbb{Q}'f''(\mathbb{Q})+p r^2  \tensor{\Gamma}{^r_\theta_\theta}f(\mathbb{Q})}{2 r  \tensor{\Gamma}{^r_\theta_\theta}f'(\mathbb{Q})}.
        \label{eq:2.partialF.2}
    \end{equation}
\end{subequations}
Since the two expressions of $F'(r)$ in Eqs.~\eqref{eq:2.partialF.1} and \eqref{eq:2.partialF.2} must be equal, we derive
\begin{equation}
    \frac{F(r)  (\tensor{\Gamma}{^r_\theta_\theta}+r) \mathbb{Q}'f''(\mathbb{Q})}{ \tensor{\Gamma}{^r_\theta_\theta}f'(\mathbb{Q})}=0,
\end{equation}
which gives rise to the three restrictive conditions, $\mathbb{Q}'=0$, $f''(\mathbb{Q})=0$,  and $\tensor{\Gamma}{^r_\theta_\theta}=-r$. The first two conditions make $f(\mathbb{Q})$ gravity reduce to GR, which has already been discussed in Ref.~\cite{d2022black}. Therefore, in order to obtain solutions beyond GR we must consider the third restrictive condition,
\begin{equation}
    \tensor{\Gamma}{^r_\theta_\theta}=-r,
\end{equation}
from which together with Eq.~\eqref{eq:option2} we derive 
\begin{equation}
    \tensor{\Gamma}{^t_r_r}=-\frac{\tensor{\Gamma}{^t_\theta_\theta}}{r^2}.
\end{equation}
Using the above two equations, we simplify Eqs.~\eqref{eq:oridif1} and \eqref{eq:oridif2} to be 
\begin{subequations}
    \begin{equation}
        \partial_r\tensor{\Gamma}{^t_\theta_\theta}=0,
    \end{equation}
    \begin{equation}
        \tensor{\Gamma}{^r_r_r}=0.
    \end{equation}
\end{subequations}
Thus we obtain the set of solutions for the connection,\footnote{Note that if we take $h=0$, the connection reduces to the solution set compatible with the Schwarzschild solution.}
\begin{equation}
    \{c,k,\tensor{\Gamma}{^t_r_r},\tensor{\Gamma}{^t_\theta_\theta},\tensor{\Gamma}{^r_r_r},\tensor{\Gamma}{^r_\theta_\theta}\}=\{0,0,-\frac{h}{r^2},h,0,-r\},\qquad h=\mathrm{const}.,
    \label{eq:2.connection}
\end{equation}
from which we can then calculate the non-metricity scalar,
\begin{equation}
    \mathbb{Q}=0.
\end{equation}
This leads to $f(\mathbb{Q})=0$, making the field equations be automatically satisfied.

As a summary to the above discussions, we have demonstrated that under Option 2 and the condition, $g_{tt}g_{rr}=\mathrm{const.}$, there are no vacuum solutions in $f(\mathbb{Q})$ gravity that go beyond GR, regardless of the specific form of $f(\mathbb{Q})$.

Because Option 1 is relatively complex, it is not addressed in Ref.~\cite{d2022black}. Although we shall not give  the set of solutions for the connection, we provide new constraints for the selection of parameters $c$ and $k$. Now let us examine Option 1. Substituting Option 1 into the vacuum metric field equations of $f(\mathbb{Q})$ gravity \cite{heisenberg2024review}, we obtain the following two equations for the $tt$ component and $rr$ components,
\begin{subequations}
\begin{multline}
\frac{\mathbb{Q}'f''(\mathbb{Q})  }{p}\left[\frac{c p (2 c-k) \tensor{\Gamma}{^r_\theta_\theta}}{F(r)}-\frac{\left(\frac{k}{2 (2 c-k)}+1\right) \left(\frac{k (k-2 c)}{2 c (2 c-k)}+2\right) F(r)}{\tensor{\Gamma}{^r_\theta_\theta}}-\frac{4 F(r)}{r}-\frac{2 p \tensor{\Gamma}{^r_\theta_\theta}}{r^2}\right]\\
-\frac{2 \left[r F'(r)+F(r)-p\right]f'(\mathbb{Q})}{p r^2}-\mathbb{Q}f'(\mathbb{Q}) +f(\mathbb{Q})=0,
        \label{eq:1.mfe1}
\end{multline}
\begin{multline}
F(r) \tensor{\Gamma}{^r_\theta_\theta} \left[ \left(2 r F'(r)+2 F(r)+p r^2 \mathbb{Q}-2 p\right)f'(\mathbb{Q})-p r^2 f(\mathbb{Q})\right]\\
-\mathbb{Q}'f''(\mathbb{Q})  \left[p  \left(c  (k-2 c)r^2-2 F(r)\right)(\tensor{\Gamma}{^r_\theta_\theta})^2+r^2 \left(\frac{k}{2 (2 c-k)}+1\right) \left(\frac{k (k-2 c)}{2 c (2 c-k)}+2\right) F^2(r)\right]=0.
        \label{eq:1.mfe2}
    \end{multline}
\end{subequations}
From Eqs.~\eqref{eq:1.mfe1} and \eqref{eq:1.mfe2}, we can solve $F'(r)$,  respectively,
\begin{subequations}
\begin{multline}
F'(r)=\frac{2 r\mathbb{Q}' f''(\mathbb{Q}) }{f'(\mathbb{Q})}
\left[\frac{p}{4}  \left(\frac{c (2 c-k)}{F(r)}-\frac{2}{r^2}\right)\tensor{\Gamma}{^r_\theta_\theta}-\frac{(k-4 c)^2 F(r)}{16 c (2 c-k) \tensor{\Gamma}{^r_\theta_\theta}}
-\frac{F(r)}{r}\right]\\
+\frac{p}{r}-\frac{F(r)}{r}-\frac{1}{2} p r \mathbb{Q}+\frac{p r f(\mathbb{Q})}{2 f'(\mathbb{Q})},
\label{eq:1.paritalF.1}
\end{multline}
\begin{multline}
F'(r)=
\frac{-4p c^2   (k-2 c)^2r^2 (\tensor{\Gamma}{^r_\theta_\theta})^2 \mathbb{Q}'f''(\mathbb{Q})}{8 c  (2 c-k)r F(r)  \tensor{\Gamma}{^r_\theta_\theta}f'(\mathbb{Q})}\\
+\frac{-4 c p (2 c-k) F(r) \tensor{\Gamma}{^r_\theta_\theta} \left[ \left(r^2 \mathbb{Q}-2\right)f'(\mathbb{Q})+2  \tensor{\Gamma}{^r_\theta_\theta} \mathbb{Q}'f''(\mathbb{Q})-r^2 f(\mathbb{Q})\right]}{8 c  (2 c-k) rF(r)  \tensor{\Gamma}{^r_\theta_\theta}f'(\mathbb{Q})}\\
+\frac{F^2(r) \left[8 c (k-2 c)  \tensor{\Gamma}{^r_\theta_\theta}f'(\mathbb{Q})+ (k-4 c)^2r^2\mathbb{Q}' f''(\mathbb{Q}) \right]}{8 c  (2 c-k) rF(r)  \tensor{\Gamma}{^r_\theta_\theta}f'(\mathbb{Q})}.
        \label{eq:1.paritalF.2}
    \end{multline}
\end{subequations}
Since the two expressions of $F'(r)$ in Eqs.~\eqref{eq:1.paritalF.1} and  \eqref{eq:1.paritalF.2} must be equal,  we deduce
\begin{equation}
 \mathbb{Q}'f''(\mathbb{Q}) \left\{4p c^2  (k-2 c)^2 r (\tensor{\Gamma}{^r_\theta_\theta})^2+F^2(r) \left[8 c (k-2 c) \tensor{\Gamma}{^r_\theta_\theta}- (k-4 c)^2r\right]\right\}=0,
\end{equation}
which gives rise to the three restrictive conditions, $\mathbb{Q}'=0$, $f''(\mathbb{Q})=0$,  and
\begin{equation}
    4p c^2  (k-2 c)^2 r (\tensor{\Gamma}{^r_\theta_\theta})^2+F^2(r) \left[8 c (k-2 c) \tensor{\Gamma}{^r_\theta_\theta}- (k-4 c)^2r\right]=0.\label{3rdsol}
\end{equation}
As analyzed for Option 2, the first two conditions make $f(\mathbb{Q})$ gravity reduce to GR, which has already been discussed in Ref.~\cite{d2022black}. Therefore, in order to find solutions beyond GR, we must consider Eq.~(\ref{3rdsol}), which gives
\begin{equation}
    \tensor{\Gamma}{^r_\theta_\theta}=\frac{F(r)\left(2 F(r)\pm\sqrt{p  (k-4 c)^2r^2+4 F^2(r)}\right)}{2cp(2 c-k)r}.
    \label{eq:1.Gamma122.pm}
\end{equation}
Again considering $\tensor{\Gamma}{^r_\theta_\theta}=-r$ in the Schwarzschild case, we take the negative branch, 
\begin{equation}
    \tensor{\Gamma}{^r_\theta_\theta}=\frac{F(r)\left(2 F(r)-\sqrt{p  (k-4 c)^2r^2+4 F^2(r)}\right)}{2cp(2 c-k)r}.
    \label{eq:1.Gamma122}
\end{equation}
Substituding Eq.~\eqref{eq:option1} and Eq.~\eqref{eq:1.Gamma122} into Eq.~\eqref{eq:oridif1}, we then obtain 
\begin{equation}
    \frac{(c-1) k p r}{c (2 c-k) F(r) \left(\sqrt{p  (k-4 c)^2r^2+4 F^2(r)}-2 F(r)\right)}=0,
\end{equation}
which leads to
\begin{equation}
    c=1 \qquad \text{or}\qquad k=0.
\end{equation}
It is a natural constraint for Option 1 to lead to solutions beyond GR: either $c=1$ or $k=0$.

We note that the term $k-4c$ appears frequently in the equations related to Option 1. However, we do not recommend simplifying these equations by setting $k=4$ when $c=1$ or setting $c=0$ when $k=0$, where the latter makes Option 1 go back to Option 2. The reason is that if we set $c=1$ and $k=4$, then Eqs.~\eqref{eq:oridif1} and \eqref{eq:oridif2} yield a solution for $F(r)$ that does not depend on the specific form of $f(\mathbb{Q})$,
\begin{equation}
    F'(r)= -\frac{3 p r^4-3 p r^2 F(r)+r^2 F^2(r)+F^3(r)}{r^3 F(r)-2 r F^2(r)}.
\end{equation}
This is a highly nonlinear first-order differential equation that cannot be solved analytically. However, the equation shows that $F(r)$ diverges as $r\to\infty$, and therefore cannot serve as the metric function for a viable black hole solution.

In this Comment, we have systematically analyzed the static and spherically symmetric vacuum solutions in $f(\mathbb{Q})$ gravity under the condition $g_{tt} g_{rr} = \mathrm{const}$. Our investigation yields two primary results. First, we have definitively shown that the simpler connection branch (Option 2) leads to a trivial $\mathbb{Q}=0$ case, which is of no physical interest in describing black holes. Second, for the more complex connection branch (Option 1), we have derived a necessary constraint ($c=1$ or $k=0$) that must be satisfied by any solution deviating from GR. While the complexity of the resulting equations under the parameter constraint makes a complete analytical solution currently elusive, this constraint itself is a significant result that was not addressed in Ref.~\cite{d2022black}. Collectively, our findings demonstrate that the condition, $g_{tt} g_{rr} = \mathrm{const.}$ (the Segre  classification $[(1,1)(11)]$), is far more restrictive in the $f(\mathbb{Q})$ framework than in GR. 


Our work proves that no exact solutions with the property $g_{tt}g_{rr} = \mathrm{const.}$ exist within the scope of the solvable Option 2 considered in the literature~\cite{d2022black} for $f(\mathbb{Q})$ gravity. This conclusion does not entirely negate the value of the approximate solutions in Ref.~\cite{d2022black}. However, our work shows that it is infeasible to introduce the assumption $g_{tt}g_{rr} = \mathrm{const.}$ in $f(\mathbb{Q})$ gravity when one solves  spherically symmetric black holes, which is an intrinsic property of the spherically symmetric black holes in $f(\mathbb{Q})$ gravity. Therefore, although the solutions in Ref.~\cite{d2022black} may still approximate real behaviors in certain aspects, our research establishes a critical constraint for future theoretical exploration: any attempt to find genuine exact spherically symmetric solutions must deviate from the condition, $g_{tt}g_{rr} = \mathrm{const}$. This provides a necessary guidance  in the study of $f(\mathbb{Q})$ gravity.

\section*{Acknowledgement}
This work was supported in part by the National Natural Science Foundation of China under Grant No.\ 12175108. Z.-X.Z.\ is also supported by the Pilot Scheme of Talent Training in Basic Sciences (Boling Class of Physics, Nankai University), Ministry of Education. L.C. was also supported by Yantai University under Grant No.\ WL22B224.

\bibliographystyle{utphys}
\bibliography{references}

\providecommand{\href}[2]{#2}\begingroup\raggedright\begin{thebibliography}{1}

\bibitem{d2022black}
F.~D'Ambrosio, S.~D.~B. Fell, L.~Heisenberg, and S.~Kuhn, ``{Black holes in
  f(Q) gravity},'' \href{http://dx.doi.org/10.1103/PhysRevD.105.024042}{{\em
  Phys. Rev. D} {\bfseries 105} no.~2, (2022) 024042},
  \href{http://arxiv.org/abs/2109.03174}{{\ttfamily arXiv:2109.03174 [gr-qc]}}.

\bibitem{Stephani:2003tm}
H.~Stephani, D.~Kramer, M.~A.~H. MacCallum, C.~Hoenselaers, and E.~Herlt,
  \href{http://dx.doi.org/10.1017/CBO9780511535185}{{\em {Exact solutions of
  Einstein's field equations}}}.
\newblock Cambridge Monographs on Mathematical Physics. Cambridge Univ. Press,
  Cambridge, 2003.

\bibitem{heisenberg2024review}
L.~Heisenberg, ``{Review on f(Q) gravity},''
  \href{http://dx.doi.org/10.1016/j.physrep.2024.02.001}{{\em Phys. Rept.}
  {\bfseries 1066} (2024) 1--78},
  \href{http://arxiv.org/abs/2309.15958}{{\ttfamily arXiv:2309.15958 [gr-qc]}}.

\end{thebibliography}\endgroup
\end{document}